\documentclass[11pt]{article}
\usepackage{iap2000,epsfig}

\def\be{\begin{equation}}
\def\ee{\end{equation}}
\def\bea{\begin{eqnarray}}
\def\eea{\end{eqnarray}}

\def \vc #1{{\mbox{\boldmath $#1$}}}

\def\nablag{{\vc \nabla}}
\def\thetag{{\vc \theta}}

\def\alphag{{\vc \alpha}}

\begin{document}
\vspace*{4cm}
\title{Cosmic Shear and Clusters of Galaxies}

\author{Y. MELLIER$^{1,2}$, L. van WAERBEKE$^{3,1}$, T. ERBEN$^4$,}
\author{P. SCHNEIDER$^5$, F. BERNARDEAU$^6$, B. JAIN$^7$, E. BERTIN$^{1,2}$}
\author{R. MAOLI$^{1,2,8}$, B. FORT$^1$, M. DANTEL-FORT$^2$}
\author{J.-C. CUILLANDRE$^{9}$, H. Mc CRACKEN$^{10}$, O. LE FEVRE$^{10}$}

\address{$^1$ Institut d'Astrophysique de Paris, 98 bis Bd Arago, 75014 Paris, France\\
$^2$ Obs. de Paris, DEMIRM, 77 av. Denfert Rochereau, 75014 Paris, France\\
$^3$CITA, 60 St. George Street, Toronto M5S 3H8, Canada\\
$^4$MPA, Karl-Schwarzscild Str. 1, 85748 Garching, Germany \\
$^5$Universitaet Bonn, Auf dem Huegel 71, 53121 Bonn, Germany\\
$^6$SPhT, CE Saclay, 91191 Gif-sur-Yvette Cedex, France\\
$^7$John Hopkins University, Dept. of Physics, Baltimore MD21218, USA\\
$^8$Dipartimento di Fisica, Universit\`a di Roma ``La Sapienza'', Italy \\
$^9$CFHT, P.O. Box 1597, Kamuela Hawaii 96743, USA\\
$^{10}$Laboratoire d'Astronomie Spatiale, 13376 Marseille Cedex 12, France\\
}

\maketitle\abstracts{The first detections of cosmic shear signal reported 
recently by 4 independent groups cover angular scales between one and 
10 arcmin. On those scales, the cosmic shear is a signature of 
  non-linear perturbations, like groups and clusters of galaxies. I present 
the results obtained by our team on CFHT and on the VLT and discuss its
impact for the analysis of cluster abundances and cosmology}

\section{Introduction}
The effects of gravitational lensing induced by large-scale structures of 
the universe
 accumulate along the lines of sight and manifest as a weak distortion on
lensed galaxies (the cosmic shear). Theoretical predictions done over the 
  last decade reveal 
that the statistical properties of cosmic shear contain
  invaluable clues on the cosmological parameters, the
 power spectrum of density fluctuation and the biasing. The cosmic 
shear is therefore an interesting way to probe the properties 
of the universe on large-scale.  \\
The amplitude of the weak cosmological distortion 
 only reaches  few percents and is contaminated by 
  artificial distortions of similar amplitude making its measurement 
  a challenging
 task.  In contrast to arc(let)s analysis, its observational signature 
 can only be recovered statistically from a large sample of galaxies 
covering several degrees of the sky.  Despite difficulties, four teams 
 challenged in order to measure the cosmic shear and get very first 
 constraints on cosmological models from this new technique. \\
Regarding the topic of this conference, the cosmic shear looks somewhat
marginal.  However, on small scales the cosmic shear signal is dominated
  by distortion produced by non-linear gravitational systems, 
like groups and clusters of galaxies. It is therefore interesting to 
  present and discuss these important results in a broader context
 including cluster formation. 
\section{Theoretical expectations}\label{section1}
Most of the mass condensations crossed by photons correspond to 
  large-scale systems with very low mass density contrasts, $\delta$.  The 
perturbation theory is therefore a valid approximation to compute the 
expected weak lensing signal.  By using the Born approximation, one
  can express the 
  deflection angle, $\alphag(\thetag,w)$,
  at angular distance $f_K(w)$ ($w$ is the radial coordinate) produced by the 
cumulative effect of the deflectors up to that distance:
\begin{equation}
\alphag(\thetag,w)= {2 \over c^2} \int_0^w {f_K(w-w') \over f_K(w)} \ 
 \nablag_{\perp} \Phi(f_K(w-w')\thetag,w') \ {\rm d} w' \ ,
\end{equation}
where $\nablag_{\perp} \Phi$ is the perpendicular gradient of the 
  Newtonian potential. Since 
the gravitational convergence, $\kappa(\thetag)$, produced by the 
 projected mass density  depends on the deflection angle, 
\begin{equation}
\kappa(\thetag)=\nablag_{\thetag}. \alphag(\thetag) \ ,
\end{equation}
one can express this ``cosmological convergence''  
 as function of the symmetric 
  components of the second derivatives of $\Phi$, that is the 
  mass density \  $\rho=\Omega_m \ \delta$.  For small perturbations, the 
first order terms of perturbative 
expansion of $\Phi$ provide a  good estimate of $\kappa$.\\
In the simple case of a single lens plane and assuming  the 
shape of the power spectrum of density fluctuations 
 is a power law (ie $P(k) \propto k^{n}$),  perturbation theory 
applied to weak cosmological lensing enables us   
 to make important statements about the use of weak lensing statistics for
cosmology:
\begin{itemize}
\item To first order, the variance of the convergence 
averaged over an angular scale $\theta$, $\langle \kappa(\theta)^2 \rangle$ 
 writes:
\begin{equation}
\langle \kappa(\theta)^2 \rangle^{1/2} \approx 0.01 \sigma_8 \Omega_m^{0.75} z_s^{0.8} 
\left({\theta \over 1^o}\right)^{-(n+2)/3} \ ,
\end{equation}
 where $\sigma_8$ is the normalization of the power spectrum, $z_s$ the 
redshift of sources.
\item Likewise, the skewness of the convergence on angular scale $\theta$, 
$s_3(\theta)$, can also be expressed as function of the same quantities:
\begin{equation}
s_3(\theta) \approx 40 \Omega_m^{-0.8} z_s^{-1.35} \ .
\end{equation}
Bernardeau et al \cite{bernardeauetal}(1997) first expressed the skewness of the convergence
for various cosmologies.  They pointed out that it can be used jointly
 with the variance to provide independently $\Omega_m$ and $\sigma_8$. 
\item Finally, the gravitational convergence can be easily related to the
 the gravitational shear, $\gamma$:
\begin{equation}
\langle \kappa(\theta)^2 \rangle=\langle \gamma(\theta)^2 \rangle \ .
\end{equation}
Since in the weak lensing regime $\gamma$ is measured directly from the 
 gravity-induced ellipticity of galaxies, it is in principle possible 
to measure the cosmic shear and to get informations about 
properties of our universe from the measurement of galaxy ellitpicities.
\end{itemize}
From an observational point of view, these statements establish the 
scientific potential of cosmic shear statistics. However, the key question
 one  needs to address is the feasibility.   Further theoretical 
studies done by Jain \& Seljak \cite{js} (1997) 
or van Waerbeke et al et al \cite{vwetal2} (2000b)  focussed on the non-linear 
regime of mass density fluctuations. On small scales, the theoretical 
expectations of first order perturbations are no-longer valid and 
the non-linear evolution of the power spectrum has to be taken 
 into account (following the Hamilton et al \cite{hamiltonetal} 1991 and 
 Peacok \& Dodds \cite{pd96} 1996
 approximations).  It turns out that, on angular scale below 10 arcmin, the 
amplitude 
of the  cosmic shear variance is twice the linear prediction and ranges between
  4 and 8 percents, according to cosmological models.  Therefore, in principle 
it should be detectable, even with 
  present-day  ground-based telescopes. 
\section{Conditions for cosmic shear detection}\label{section2}
The design of the survey and the accuracy of PSF-anisotropy
 correction control the successful outcome of any cosmic shear survey.\\
 The size of the survey is critical. In order to 
 constrain cosmological scenarios, it must at least cover an angular 
scale which  
   provides a cosmic shear signal-to-noise ratio higher than 3 for 
any model one could reasonably expect. Van Waerbeke et al \cite{vwetal99} (1999) addressed 
 this issue and used extensive simulations 
to infer the best strategy. Their main results are listed in 
Table \ref{sizesurvey}. 
 It shows that one needs to survey about one deg$^2$ to get a significant 
  estimate of the variance of $\kappa$, whereas about 10 deg$^2$ are needed 
  to probe
  its skewness.  
These
predictions may be somewhat pessimistics and the size of the survey 
could be lower than the van Waerbeke et al simulations, because 
 the enhancement of the shear produced by non-linear effects
 on small scales were not taken into account.
 In practice, since many ongoing surveys do have already 
covered one
deg$^2$, at least the variance can be measured easily right now.

\begin{table}[t]
\caption{\label{sizesurvey}Expected signal-to-noise ratio on the measurement of the variance and the skewness of the convergence for two extreme realistic 
  cosmological models.  In the first
column, the size of the field if view (FOV) is given. The signal-to-noise
 ratio is computed from the simulations done by 
van Waerbeke et al  (1999). The top line of this table describes 
shortly some details of the analysis.  The redshift of the sources and
the galaxy number density correspond to a typical 2-hours exposure on a
4-meter telescope.}
\begin{center}
\begin{tabular}{|l|c|c||c|c|}
\hline
\multicolumn{5}{|c|}{{\large {$z_s=1$, Top Hat
Filter , $n=30$ gal.arcmin$^{-2}$}}} \\ \hline
\hline
 {FOV} &
\multicolumn{2}{|c|}{ {S/N Variance}} & 
\multicolumn{2}{|c|}{ {S/N Skewness}} \\
\cline {2-5}
{\large { (deg.$\times$deg.) }}&  {$\Omega_m=1$}  &
 {$\Omega_m=0.3$}  &
 {$\Omega_m=1$}  &
 {$\Omega_m=0.3$}  \\ \hline
{\large {1.25$\times$1.25}}  & {\large {7} } & {\large {5}}
&{\large { 1.7} }
 & {\large { 2} }\\ \hline
{\large {2.5$\times$2.5}}  & {\large {11} } & {\large {10}}
&{\large { 2.9} }
 & {\large { 4} }\\ \hline
{\large {5$\times$5}}  & {\large {20} } & {\large {20}}
&{\large { 5} }
 & {\large { 8} }\\ \hline
{\large {10$\times$10}}  & {\large {35} } & {\large {42}}
&{\large {8} }
 & {\large { 17} }\\ \hline
\end{tabular}
\end{center}
\end{table}
 
\begin{figure}
\psfig{figure=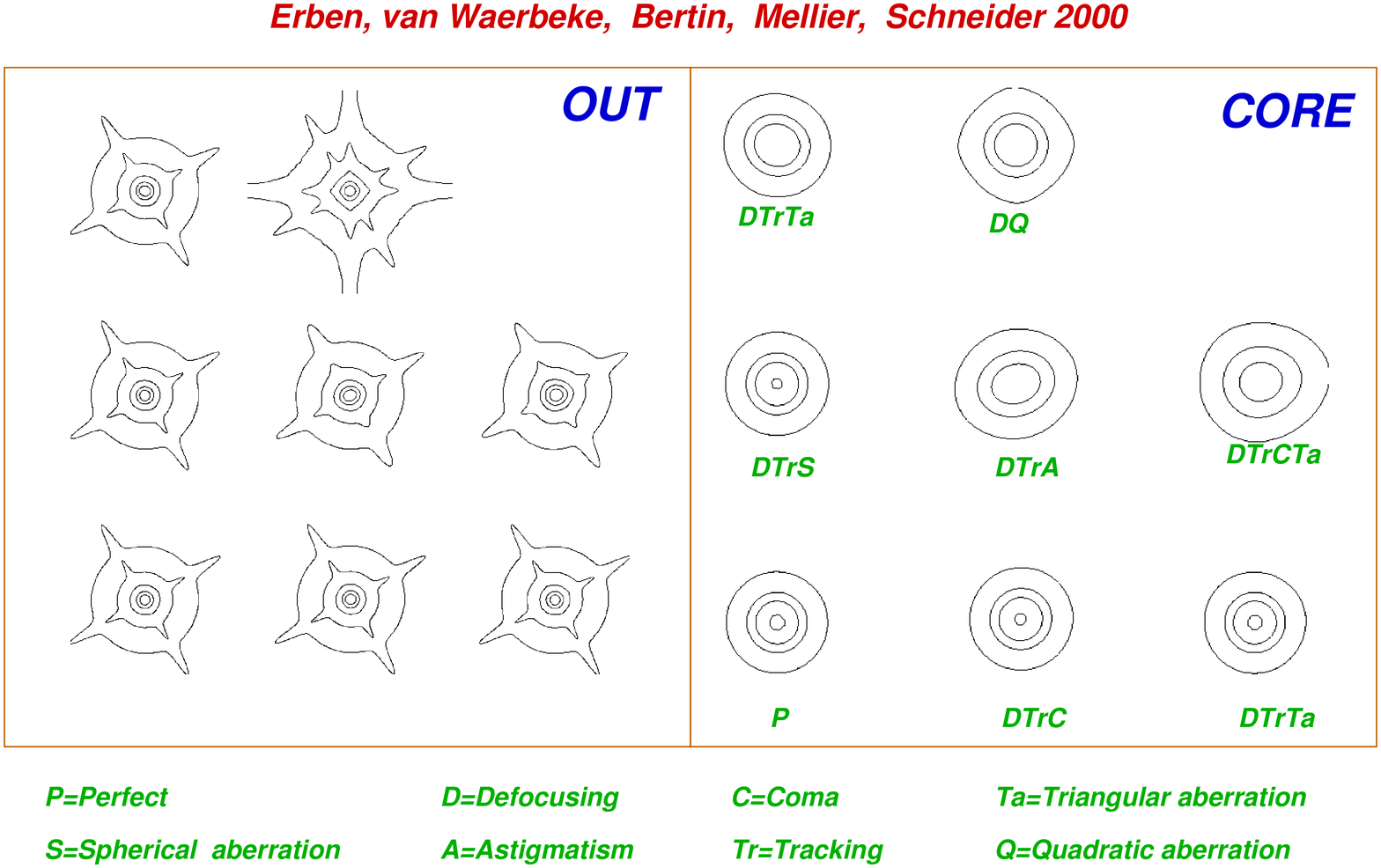,height=10cm}
\caption{The various optical distortions simulated by 
 Erben et al (2000)
in order to evaluate the limiting performances of KSB. The left panel 
shows the external shapes of the PSF for each set of parameters listed 
below the stars. On the right panel, the central part of the PSF has been
 zoomed in order to show details of internal shapes. For all these models
 the shear can be recovered down to 0.01 with 10\% relative accuracy.
\label{erbenfig}}
\end{figure}

The feasibility of the detection of cosmic shear critically depends 
on the accuracy of the PSF anisotropy correction.  These anisotropies
result of atmospheric distortion (atmospheric dispersion), 
  technical or optical problems (bad telescope tracking, 
telescope flexures, charge transfer inefficiency, optical distortions)
 or astronomical artifacts (saturated stars or diffusion halos produced 
  by multiple reflexions
 from bright objects).  The correction of PSF anisotropy in weak lensing
analysis has been addressed by many groups over the last years 
(see Mellier \cite{mellier99} 1999 and Bartelmann \& Schneider \cite{bs00} 2000 for reviews). The 
technique developed by Kaiser, Squires \& Broadhurst \cite{ksb} (1996, hereafter KSB) 
  is now widely used and recent extensive simulations 
 show that, by using the KSB correction,  weak lensing signals
   can be recovered without biases
 (Hoekstra et al \cite{hoekstraetal98} 1998, Erben et al \cite{erbenetal00}
2000, Bacon et al \cite{baconetal2} 2000).  Erben 
et al (2000) have processed thousands of simulated images
  containing a wide variety of PSFs with most of the optical
  distortions one could expect. 
 They are listed in Fig. \ref{erbenfig}.  They conclude that any 
shear amplitude in the range $0.012<|\gamma|< 0.300$ can be recovered with a 
10\% relative accuracy.  This important result implies that the cosmic
shear should be detected even in usual ground-based images, provided 
the seeing is smaller than 1 arcsecond. 

\section{First detection of cosmic shear}\label{section3}
\begin{figure}
\psfig{figure=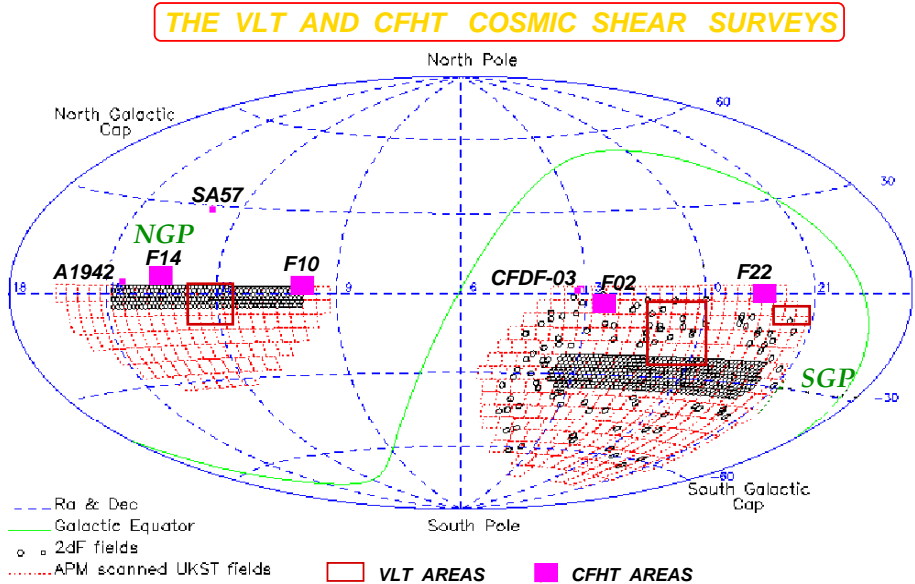,height=10cm}
\caption{Positions of the areas covered by the CFHT and VLT cosmic shear 
surveys. The filled areas are the targets covered at CFHT.  The open rectangles 
 delineate the areas inside which the 45 VLT targets were selected.  
\label{oursurvey}}
\end{figure}

\begin{figure}
\centerline{
\psfig{figure=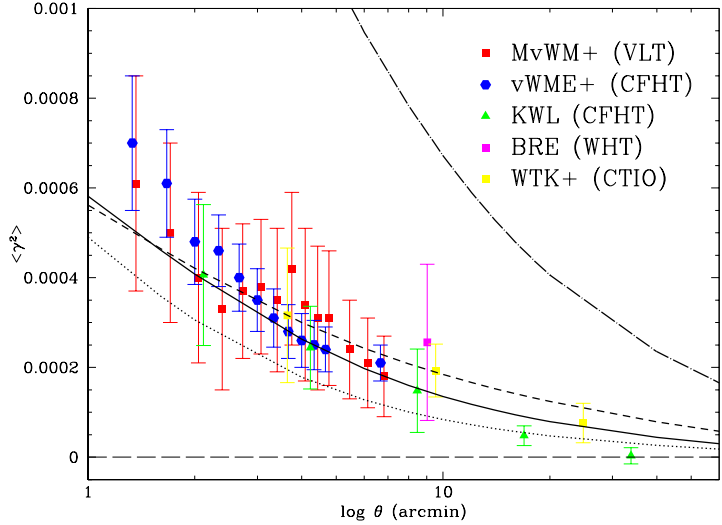,angle=270,width=10cm}}
\caption{\label{cosmicshear}{\it .
The recent results of cosmic shear measurements. The  works
 referred to as Maoli et al 2000 (MvWM+), 
   van Waerbeke et al 2000 (vWME+), 
 Kaiser et al 2000 (KWL), Bacon et al 2000 (BRE) and Wittman et al
 2000 (WTK+). 
Some predictions of cosmological models are also plotted, assuming 
sources at $z_{eff}=1$ and using the non-linear evolution of 
power spectrum according to the coefficients given by Peacock (1999). The solid line corresponds to $\lambda$CDM, 
with $\Omega_{m}=0.3$, $\lambda=0.7$, $\Gamma=0.21$; the dot-dashed line
 to COBE-normalized SCDM; the dashed line to 
cluster-normalized SCDM and the dotted line to 
cluster-normalized Open CDM with $\Omega_{m}=0.3$. 
\label{fig:results}
}
}
\end{figure}

Thanks to the optimistic conclusions on the feasibility of a cosmic  
 shear survey and 
reliability of PSF corrections, four teams 
 carried out a wide field survey of galaxies in order to detect cosmic 
shear signatures on scales ranging from one to ten arcminutes: 
\begin{enumerate}
\item  Van Waerbeke et al \cite{vwmeetal} (2000) 
covered 1.7 deg$^2$ up to $I=24$ over 5 uncorrelated fields obtained at 
  CFHT, 
\item Bacon et al \cite{baconetal1} (2000) investigated 14 fields up to
  $R=24$ obtained 
 at WHT over 0.5 deg$^2$, 
\item  Wittman et al (2000) observed 
 1.5 deg$^2$ up to $R=26$ over 3 uncorrelated fields at CTIO, 
\item  Kaiser et al \cite{kaiseretal} (2000) analysed 6 fields up to $I=24$
  at CFHT, covering 1. deg$^2$, 
  and finally
\item  Maoli et al \cite{mvwmetal} (2000)
  used 45 VLT fields covering 0.5 deg$^2$ up to $I=24$.  
\end{enumerate}
The teams used different instruments, observed different fields 
of view and used different 
 techniques to analyze the data and correct for the PSF anisotropy
   which enables to check the reliability and 
the confidence level of the results. This is indeed a chance to get 
  these works published almost simultaneously.\\
Our CFHT and VLT surveys are reported in van Waerbeke et al 
(2000) and Maoli et al (2000) respectively (see Fig. \ref{oursurvey}).   Thanks to these independent 
  data set, we were able to cross-check 
our results and investigate the reliability of our 
 corrections of  systematics. Details can be found
  in van Waerbeke et al (2000).  The 45 VLT are of special interest 
because the targets are spread over more than 1000 squares degrees, each 
of them being separated from the others by at least 5 degrees.  These 
uncorrelated fields provide a direct measurement of the cosmic variance, 
without need of simulations. \\
The four teams announced a detection of cosmic shear signal during 
 the  first semester 2000.
The results are summarized in Fig. \ref{cosmicshear}.  The most striking 
  feature on this plot is  
  the remarkable similarity of the signatures in the range 1' to 10' . This is 
  obviously a crucial point which make us confident that the 
  detection and  measurements are reliable and  robust, despite
   concerns about systematics.  
%
\section{What do we learn about cosmology and clusters?}\label{section4}
The comparison of the measurements with some typical cosmological 
models displayed in Fig. \ref{cosmicshear} (the non-linear evolution of
the power spectrum is computed with the coeeficient given in Peacock
\cite{pik} 1999, which seems to provide lower amplitudes of the
 variance of the shear
 than previous coefficients) leads to the following 
conclusions:
\begin{itemize} 
\item Although the points as function of angular scale  are correlated, the
simultaneous use of independent data provided by the five groups 
   permits to reject 
 some models with a very high significance.  In particular,
  the SCDM COBE-normalized model is ruled out to at least a 5-$\sigma$ level. 
\item In contrast, the popular $\lambda$-CDM cluster-normalized model  
perfectly fit the data, but this is not the only one. The cluster-normalized
 SCDM as well as a cluster-normalized open universe
 ($\Omega_m=0.3$) are also compatible with the observations. This illustrates 
that error bars are still too large and also that the variance of the shear 
  is not enough to break the degeneracy ($\sigma_8,\Omega_m$).  This 
is only once the skewness of the convergence will be measured that 
we will be in much better position to constrain cosmological
  scenarios.
\end{itemize} 
The depth of these surveys corresponds to sources at redshift of about 
$ z \approx 0.8-1$. The typical efficiency function, which describes the
 lensing strength of the lenses as function of the redshift
distributions of the lenses and the sources,  
  should therefore peaks at redshift $z \approx 0.4$.
 On angular scales between 1' and 10', since the non-linear structures dominate 
  the signal, most of the cosmic shear is produced by structures having
physical sizes of about $0.2 - 1.0 $ $h^{-1}$ Mpc.  The cosmic shear surveys
  are therefore mainly probes of weak cosmological lensing produced by
 clusters of galaxies and compact groups.  \\
The constraints provided by cosmic shear are formally similar to 
those from cluster abundances obtained from 
  counts of clusters in optical or X-ray surveys (Eke et al \cite{Eke96} 1996, 
 Eke et al \cite{Eke98} 1998;
 Bridle et al \cite{Bridle99} 1999). Depending on the angular scales, 
 the variance of the cosmological convergence writes:
\begin{equation}
\langle \gamma^2\left(\theta<10'\right) \rangle^{0.5} \propto
 \sigma_8^{^{1.3}} \Omega_m^{^{0.65}}  \ \ , \ \ 
\langle \gamma^2\left(\theta>10'\right) \rangle^{0.5}
\propto \sigma_{_8}  \Omega_m^{^{0.8}} \ ;
\end{equation}
whereas, for  cluster abundances the constraints have formally the following
 dependences:
\begin{equation}
\sigma_{_8}  \Omega_m^{^{\approx 0.55}} \approx 0.6  \ .
\end{equation}
The cosmic shear has the advantage of being a direct measurement of the 
lensing 
effects produced by dark matter.  In contrast, the cluster abundances
measures the fraction of massive clusters from the light 
distribution, which implies, either empirical relation 
between light and mass (like emissivity-temperature relation), or 
  assumptions 
  of the geometry and the physical state of the baryonic and
non-baryonic components. \\
More interesting,  in principle, one could break the degeneracy by using  
  cluster abundances and cosmic shear as independent 
 data sets.  Up to now, the uncertainties of cosmic shear signal as well
  as cluster abundances are still too large.  But this looks like  
 a very promising approach, which 
   will be feasible soon by using the complete and well-defined X-ray 
samples provided by XMM, the cluster abundances from the 
VIRMOS-optically selected sample or even the future SZ-surveys.

\section{Future prospects}\label{section5}
The first detection of cosmic shear by various groups puts solid grounds
  on the weak lensing approach as a biased-free cosmological tool. 
 We expect that ongoing surveys will provide soon similar 
  constraints as those from CMB or SNIa projects.  Within the next five 
 years, 
 wide field cosmic shear surveys will produce the first measurements 
 of the variance,  the skewness of the convergence as well as 
 alternative statistics - like the peak statistics, the genus or
  the foreground-background correlations - 
 and the properties of the  power spectrum of mass density fluctuations up
  to degree scales.  The full mass range of structures will 
be available then, with  very accurate cosmological constrains
 (van Waerbeke et al 1999). \\ 
In parallel, we still have to analyze carefully the effects of
 systematics and the validity of some approximations which are 
  the main pillars of cosmic shear analysis.
 Up to now, the investigations carried out in the realm of 
  observations, numerical simulations and perturbation theory lead to 
  detailed studies on the followings issues:
\begin{itemize}
\item {\sl Validity of the Born approximation:}
 The effects of mass density fluctuations 
  on the deformation of the ray bundles are computed assuming that the 
 the deformation can be computed along the unperturbed geodesic. 
\item {\sl Lens coupling}:
 When ray bundles eventually cross two
lenses, the convergence produced by a lens depends on the 
  lensing effects produced by other structures.
 Hence, the 
 magnification matrix is not simply the sum of the 
two convergences but also contains  additional coupling terms. 
 \item {\sl Redshift of sources}:
Both the variance and the
skewness of the convergence strongly depend on the redshift 
  distribution of the lensed sources ($\approx z^{0.7-1.4}$).  
\item {\sl Source clustering}:
Due to galaxy clustering, the amplitude of the gravitational shear 
may strongly vary from one line-of-sight to another. 
 The average redshift distribution of the sources can therefore
be biased by the galaxies located within the massive structure,  
which bias the value of the convergence in a similar way.
\item {\sl Intrinsic correlated polarization of galaxies}
If the intrinsic orientations of galaxies  
 is not randomly distributed, their  
coherent alignment may correlate to the geometry of large
scale structures in which they are embedded. If this is the case, the coherent 
alignment produced 
by weak lensing will be contaminated by the intrinsic 
 alignment of the galaxies and a mass reconstruction 
 based on the shear pattern will be strongly contaminated by spurious weak
lensing signal.  Recent analyses carried out by Croft \& Metzler
\cite{croft} (2000) and Heavens et al. \cite{heavens} (2000) conclude 
that on scale smaller than 10 arcminutes the intrinsic correlation 
should not contaminate the weak lensing signal, provided the survey
 is deep enough in order to probe distant lensed galaxies.  However, 
 for shallow survey the conclusions are unclear.
\end{itemize}
In general, it turns out that most of these issues 
do not have a major negative impact 
on the cosmic shear. The recent ray tracing analysis done  by Hamana et al
\cite{hamana} (2000) shows that neglecting both the full ray tracing and
  the lens-lens coupling have negligible effects on the results, which 
is confirmed by semi-analytic computation (Bernardeau et al
\cite{bernardeauetal} (1997), van Waerbeke et al \cite{vwetal2} (2000b)).
   The most critical seems to be the source clustering which
  could change the signal by 20\% (Thion et al \cite{thion} 2000; 
 Hamana et al in preparation). Fortunately, 
  this can be reduced to 
 1\% if one uses sources within a narrow redshift range. 

The detection of cosmic shear and the demonstration that reliable 
results come out from it are  one of the most important 
results in observational cosmology this year.  Indeed, we hope it may  
  put cosmic shear projects at the same level  
 as CMB  and SNIa experiments. Among those, cosmic shear is the
only one which probes directly the distribution of dark matter. 
\\



\section*{Acknowledgements} We thank T. Hamana and A. Thion for fruitful
discussions. This work was supported by the TMR Network ``Gravitational
Lensing: New Constraints on Cosmology and the Distribution of Dark
Matter'' of the EC under contract No. ERBFMRX-CT97-0172.

\end{document}